# Synchronization of distant optical clocks at the femtosecond level

Jean-Daniel Deschênes[*], Laura C. Sinclair, Fabrizio R. Giorgetta, William C. Swann, Esther Baumann, Hugo Bergeron, Michael Cermak, Ian Coddington, Nathan R. Newbury[*]

*National Institute of Standards and Technology, 325 Broadway, Boulder, Colorado 80305*
[*] *jean-daniel.deschenes@gel.ulaval.ca, nathan.newbury@nist.gov*

The use of optical clocks/oscillators in future ultra-precise navigation, gravitational sensing, coherent arrays, and relativity experiments will require time comparison and synchronization over terrestrial or satellite free-space links. Here we demonstrate full unambiguous synchronization of two optical timescales across a free-space link. The time deviation between synchronized timescales is below 1 fs over durations from 0.1 s to 6500 s, despite atmospheric turbulence and kilometer-scale path length variations. Over several days, the time wander is 40 fs peak-to-peak. Our approach relies on the two-way reciprocity of a single-spatial-mode optical link, valid to below 225 attoseconds across a turbulent 4-km path. This femtosecond level of time-frequency transfer should enable optical networks using state-of-the-art optical clocks/oscillators.

**Subject Areas** Optics, Atomic and Molecular Physics, Optoelectronics

## I. Introduction

Optical clocks reach absolute accuracies approaching $10^{-18}$ [1–5] and optical oscillators (e.g. cavity-stabilized lasers) can provide sub-femtosecond timing stability over seconds [6–10]. A physical network of optical timescales, derived from these clocks, could enable dramatic improvements in precision navigation and timing [5,11,12], phased sensor arrays, tests of special and general relativity [5,13–16], clock-based geodesy [5,17–19], and even future searches for dark matter [20]. In these applications, the local optical timescale would either be compared against or synchronized to a distant timescale via terrestrial or satellite free-space links. In particular, synchronization of distant optical timescales could enable future high-precision navigation/timing networks, e.g. an optically-based GPS, by synchronizing compact optical oscillators to a few, more complex and larger master optical atomic clocks. In implementing such networks, optically-based time-frequency transfer techniques will be needed because rf-based free-space time transfer is limited to 10 ps to 100 ps accuracy and ~1 ps stability, a hundred to a thousand times worse than optical oscillators [21–23]. The highest performance rf system to date is planned for the ACES mission and will support 300 fs timing stability at 300 s integration and < 6 ps timing stability over days from ground to space [24,25]. Clearly, optical clocks/oscillators with femtosecond precision will eventually require analogous optical time-frequency transfer techniques to realize their full potential. Optical fiber-based techniques achieve frequency transfer at $10^{-18}$ fractional stability over 1840 km [26,27], time transfer at sub-picosecond stability over up to a thousand kilometers [28,29], and sub-femtosecond stability over several kilometers [30–33]. These fiber links are appropriate for connecting national laboratories but to support the broader applications of clock networks, free-space links will be essential [34–36].

The challenge with comparing and synchronizing the time between distant clocks arises from the finite speed-of-light. A direct comparison of their time inevitably includes an unknown and variable path delay in transmitting one clock signal to the other. In two-way rf comparisons, this problem is overcome by transmitting the time signals between clocks in each direction. Subtraction of the measured arrival times then yields the clock time offsets independent of the path delay - provided the path is reciprocal with equal delay both directions. Ref. [35] introduced an analogous two-way time-frequency transfer approach in the optical domain based on frequency combs. In that work, the goal was to enable frequency comparisons between remote clocks after post-processing, although the demonstration made use of a common optical oscillator. Here, we pursue the much more challenging problem of two-way *time* comparison between two distant optical timescales and, with active real-time feedback, their time synchronization. The ability to not just compare but to synchronize two distant clocks at the femtosecond level over the air can be a powerful tool but has significant complexity as it requires real-time measurements of the absolute time offset between clocks, real-time communication between sites, and real-time adjustment of the synchronized clock, all with femtosecond level precision. Moreover, to achieve femtosecond time synchronization, this two-way approach must cancel variations in the path length between the distant sites to below 300 nm despite kilometer or longer paths through turbulent air.

Here, we show optical two-way time transfer can indeed compare and synchronize two optical timescales to the femtosecond level and at an update rate of 0.5 milliseconds. The basic setup is sketched in Fig. 1a. In this work, we construct two optical timescales based on independent cavity-stabilized lasers (i.e. optical oscillators). We show these two timescales can be time synchronized with sub-femtosecond stability from 0.1 s to 6500 s (See Fig. 1b), dropping as low as 225 attoseconds at 10 seconds averaging. Over two days, the long-term wander of the time offset is 40 fs peak-to-peak which is attributed to temperature induced variations in the non-reciprocal fiber paths associated with the out-of-loop time verification and the coherent transceivers. This femtosecond-level performance is reached despite strong turbulence-induced fading and piston noise [37,38], variations of hundreds of picoseconds in the path delay from turbulence and weather, temporary misalignments of the link, and intentional variations of the path length from 1 m to 4 km. At these levels, a network of optical clocks/oscillators will have a sensitivity a thousand times superior to analogous rf-based systems for timing, navigation and sensing.

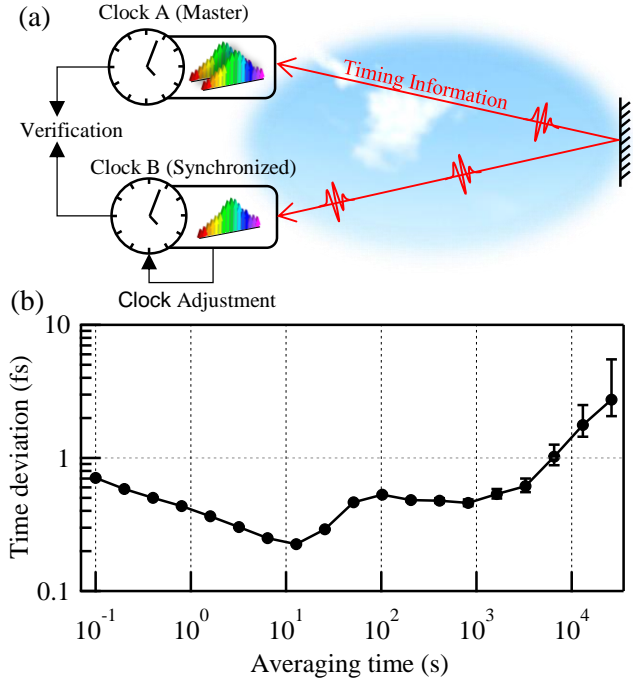

FIG 1. (a) General synchronization concept. Time information is transmitted between sites across a turbulent air path. Real-time feedback is applied to the clock at site B to synchronize it with the clock at site A. A folded optical path allows for verification of the synchronization by a direct "out-of-loop" measurement. (b) Measured timing deviation, or precision, between the time outputs while synchronized across a 4 km link, based on data acquired over two-days as described in Section IV.

## II. Synchronization between two distant optical timescales using two-way time transfer

A first requirement is to create an optical timescale at each site. The name notwithstanding, state-of-the-art atomic optical clocks are operated as frequency standards; they output an optical frequency from a laser stabilized to an optical cavity and atomic transition. Therefore, atomic optical clocks are compared by their frequency ratios, typically via a frequency comb. To create a timescale, we phase lock a self-referenced frequency comb to the cavity-stabilized laser. The optical pulses output by the frequency comb are then analogous to the "ticks" of a conventional clock. To generate a local time, we then define a reference plane and label the comb's optical pulses according to their arrival at the reference plane. In other words, a local controller tracks the pulse number and converts this to a corresponding time using the assumed underlying cavity-stabilized laser frequency. This extension to a timescale does require a phase-continuous connection between the comb and the cavity stabilized laser. More generally, this laser could be stabilized to an atomic transition to provide an absolute timescale at a single master site or at both sites for relativity experiments.

We first review the simplest implementation of two-way transfer, before discussing the modified optical two-way transfer demonstrated here. Consider two clocks at separate sites A and B. Suppose site A transmits a pulse at its zero time to site B. Its measured arrival time according to site B's clock is $\Delta \tau_{A \to B} = T_{link} + \Delta T_{AB}$ where $T_{link}$ is the path delay and $\Delta T_{AB}$ is the time offset between the sites. Simultaneously, site B transmits a pulse at its assumed zero time in the opposite direction to site A, where its arrival time is $\Delta \tau_{B \to A} = T_{link} - \Delta T_{AB}$. Subtraction of these two arrival times yields the clock time offset, $\Delta T_{AB} = \frac{1}{2}(\Delta \tau_{A \to B} - \Delta \tau_{B \to A})$ that must be adjusted by a calibration constant, $\tau_{cal}$ to account for time offsets in the transceivers. Summation of the two signals provides $T_{link}$ and therefore the distance between sites given the speed of light.

The optical clock output is a train of very short optical pulses with femtosecond residual timing noise. However, we cannot implement the simple two-way protocol discussed above with direct photodetection of the pulse arrival time, as this immediately introduces picosecond-level uncertainty due to the limited photodetector response time and low light levels (for any reasonable link loss). Instead we implement linear optical sampling between frequency combs with offset repetition rates [35]. The overall setup is illustrated in Fig. 2.

Linear optical sampling requires the transmission of comb pulse trains between sites that differ in repetition rate. On the other hand, synchronization requires optical timescales (comb pulse train) at each site that operate at the same repetition rate. Therefore, we require three combs: a comb at each site with a repetition rate $f_r$, and a third transfer comb at site A, with a repetition rate $f_r+\Delta f_r$. (This configuration supports not only comparison but the synchronization of site B to the master site A.) The relative evolving time offsets between pulses from the three frequency combs are measured through linear optical sampling via the three balanced photodetectors in Fig. 2 to retrieve: (i) the time offset between the comb pulses from Site B and the transfer comb pulses at site A, $\Delta \tau_{B \to X}$, (ii) their analogous time offset at site B, $\Delta \tau_{X \to B}$, and (iii) the time offset between the transfer pulses and the comb pulses at site A, $\Delta \tau_{X \to A}$. As outlined in Appendix B, we can then derive a "master synchronization equation" – the analogy of the simple two-way time transfer equation given earlier – for the time offset between site A and site B as,

$$\Delta T_{AB} = \frac{1}{2}(\Delta \tau_{B \to X} - \Delta \tau_{X \to B}) - \Delta \tau_{A \to X} + \tau_{cal} - \left(\frac{\Delta f_r}{2 f_r}\right)(T_{link} + \Delta t_{ADC}) + \frac{\Delta n}{2 f_r} \quad (1)$$

where $T_{link}$ is the time-of-flight across the link, $\Delta t_{ADC}$ is the time offset in the analog-to-digital converters (ADCs) at the two sites, $\Delta n$ is an integer related to the pulse labelling, and $\tau_{cal}$ is a calibration offset related to transceiver delays and the selected location of the reference planes. Eq. (1) yields the time offset $\Delta T_{AB}$ between clocks every update period of $1/\Delta f_r$. To synchronize site B to the master site A, this value is input to a phase-locked loop that controls the timing at site B thereby driving $\Delta T_{AB}$ to zero with a time constant set by the phase locked loop at some multiple of the update time $1/\Delta f_r$.

The first three terms of Eq. (1) comprise a generalized two-way time transfer expression. As derived in Appendix B, there are the two additional terms, one proportional to $\Delta f_r$ and one proportional to $\Delta n$. The latter simply accounts for the $1/(2f_r)$ ambiguity in the pulse labeling. The former is a small correction accounting for the mismatch in repetition rates between the transfer comb's pulse train and the optical timescales. This mismatch is necessary for the linear optical sampling, but leads to an incomplete cancellation in the two-way comparison of both the path delay and the relative timing of the digitizers used with the photodetectors. The term is small since it is proportional to $\Delta f_r/(2f_r) \sim 1/200,000$ but its inclusion is needed for correct time comparison and synchronization.

The frequency-comb-based measurements cannot provide a value for these last two terms. Instead, we require a "coarse" two-way time transfer measurement that measures $\Delta t_{ADC}$ and $T_{link}$ without ambiguity and with an uncertainty below $2f_r/\Delta f_r \times 1$ fs for femtosecond uncertainty in $\Delta T_{AB}$. It must also measure $\Delta t_{ADC}$ to below $1/(2f_r)$ to resolve the integer $\Delta n$. In our system, the uncertainty of this "coarser" two-way time transfer needs to be below 200 ps, which is well within the capabilities of rf-based systems.

Finally, calculation of Eq. (1) requires information exchange between sites, which in turn requires rapid, real-time communication. Optical communication across a free-space link is well known to suffer from dropouts due to turbulence. Here, however, that problem is moot, as the optical communication channel uses the same single mode spatial link as the comb light. Any turbulence-induced dropouts are correlated and therefore communication is available whenever the timing information is available.

### III. Experimental implementation

Figure 2 shows a high-level view of the physical system. Sites A and B are connected via a free-space single-spatial-mode optical link covering up to 4 km. This link is folded by use of two plane mirrors so that sites A and B are physically adjacent, enabling synchronization verification via an out-of-loop measurement of the actual time offset, $\Delta T$, independent of the "in-loop" predicted time offset $\Delta T_{AB}$.

The full system includes two cavity-stabilized lasers, two Doppler cancelled links that carry the light from the cavity-stabilized lasers to the frequency combs, three self-referenced combs, a coherent communication link, coarse two-way phase modulated time transfer, a feedback system to the site B timescale (the synchronized site), field-programmable gate array (FPGA) controllers, and two free-space terminals. We discuss some of the salient details below and in Appendix C.

The cavity-stabilized lasers for both sites are located in an environmentally stable laboratory that is 200 m from the main transceivers. A commercial cw fiber laser is locked to an optical cavity with a ~1 Hz linewidth and a typical environmentally-induced frequency drift ranging from 0 Hz/s to 10 Hz/s. The frequency of the cavity-stabilized laser is 195.297,562 THz for site A and 195.297,364 THz for site B. Two separate Doppler-cancelled fiber links transport these frequencies to the comb-based transceivers located in a rooftop laboratory. The phase-lock of the cavity-stabilized lasers and the Doppler cancelled links are monitored during synchronization to ensure that no phase slips occur.

The three frequency combs are each self-referenced with FGPA-based digital control and can operate for days without phase slips [39]. The 972,920th mode of site A's frequency comb is phase-locked to its cavity-stabilized laser to yield a repetition rate of 200.733,423 MHz. The transfer comb is phase-locked to the same cavity-stabilized laser to yield a repetition rate that differs by $\Delta f_r = 2.27$ kHz. At site B, the 972,919th comb mode is locked to the second cavity-stabilized laser, yielding a repetition rate very close to the comb at site A. The rf offset of this phase-lock is adjusted to maintain synchronization. For loop stability considerations, the bandwidth of this feedback should be below $\sim \Delta f_r / 4 = 500$ Hz. Here, however, based on the free-running noise of the cavity-stabilized laser and measurement noise level on $\Delta T_{AB}$, a 10 Hz feedback bandwidth minimizes the residual jitter. The combs, as well as the heterodyne detection modules, are enclosed in small temperature controlled aluminum boxes within a larger transceiver box, which is loosely temperature controlled. A 16 nm wide (2 THz) section of the comb spectrum, centered at 1555 nm, is transmitted over the link with a power at the transmit aperture of 2.5 mW.

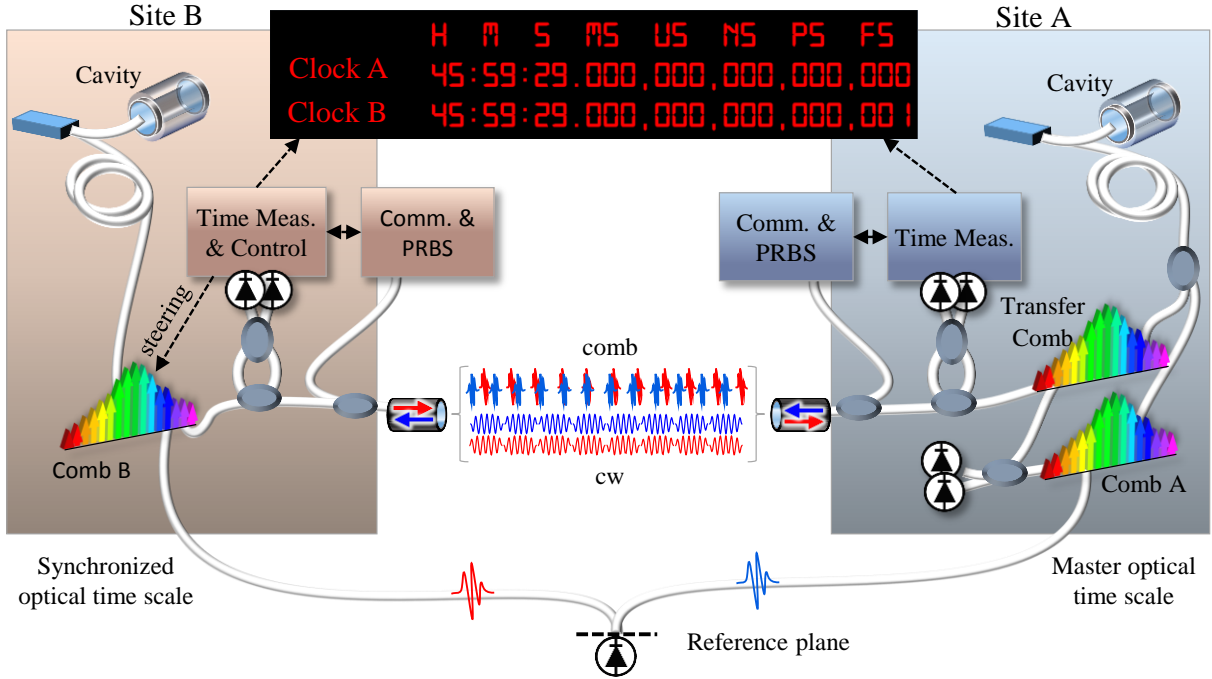

FIG 2. (Color, two column) Two optical timescale, at site A and site B, each comprise a cavity-stabilized laser, Doppler-cancelled link, and self-referenced frequency comb. The clock "time" is defined by the arrival of the comb's pulses at the reference plane. (The controller tracks the time label associated with each pulse.) Two-way optical exchange over the free-space link allows comparison, and further synchronization, of the two timesclaes so that their pulses arrive at the reference plane at a common time, as shown in the upper display. An entire unambiguous measurement of the time offset requires only 0.5 ms; therefore, the system operates robustly over the intermittent free-space link.

We implement the "coarse" two-way time transfer, needed to establish the rightmost terms in Eq. (1), as in rf-based two-way time frequency transfer [40], except that the timing signal is carried by phase modulation of an optical carrier (cw laser) with a pseudorandom binary sequence (PRBS), as described in Appendix C. The optical carrier traverses the same single-mode spatial optical link as the two-way comb light and therefore measures the same path delay over the air. The modulated cw light does have a different total path delay than the combs because of non-common mode fiber optic paths in the transceivers, but these are effectively included in the calibration of $\tau_{cal}$ and their variations are suppressed by $\Delta f_r / (2 f_r) \approx 1/200,000$ in Eq. (1). As implemented here, this coarse two-way time transfer determines the unambiguous value of $T_{link}$ (and $\Delta t_{ADC}$) with an uncertainty of 40 ps, which is more than sufficient to support femtosecond synchronization via Eq. (1). The real-time optical communications link is implemented across the free-space link using the same hardware as the coarse two-way time transfer, as discussed in Appendix C. The communication is interposed between the coarse two-way time measurements within a single $1/\Delta f_r$ interval so that Eq. (1) is updated at $1/\Delta f_r = 0.5$ ms intervals. Finally, the modulated cw light and comb light are combined within the same single mode fiber and launched via free-space optical terminals with tip/tilt control to compensate for turbulence-induced beam wander.

### IV. Results

As shown in Fig. 1, we verify the time synchronization by direct "out-of-loop" measurements of the time offset, $\Delta T$, that are completely independent of the calculated value, $\Delta T_{AB}$.

The most sensitive measurement of $\Delta T$ is achieved by heterodyne detection between the two optical timescale outputs – i.e. the 200 MHz pulse trains from the combs – at a common reference plane. To do this, the carrier-envelope offset frequency of the frequency comb at Site B is purposefully offset relative to the comb at Site A by 1 MHz. In this case, the heterodyne signal of comb pulses overlapping in time at the reference plane appears at 1 MHz with an amplitude that depends on their time offset, as illustrated in Fig. 3a. The response is measured with a shorted link and $\tau_{cal}$ is selected so that a nominal zero time offset falls in the linear response region, i.e. the blue dot in Fig. 3a. The system is then operated over the link and the scaled demodulated amplitude provides a direct measurement of $\Delta T$ at 0.5 ms intervals, as shown in Fig. 3b. Over the one hour interval, the full standard deviation is 2.4 femtoseconds. The next section provides similar data over a longer time period and for varying path lengths.

This heterodyne detection between the 200 MHz pulse trains does not verify that the timescales are unambiguously synchronized, i.e. that there are no 5-ns slips. Section IV.C provides data on comparison of an optical pulse-per-second output through direct photodetection. It also compares synchronous 1 Hz pulse bursts through direct detection of spatial interference fringes between the optical pulses. For the latter, we are observing optical spatial interference between the ~100 fs optical pulses of two sources that are connected only via a 4 km free space link.

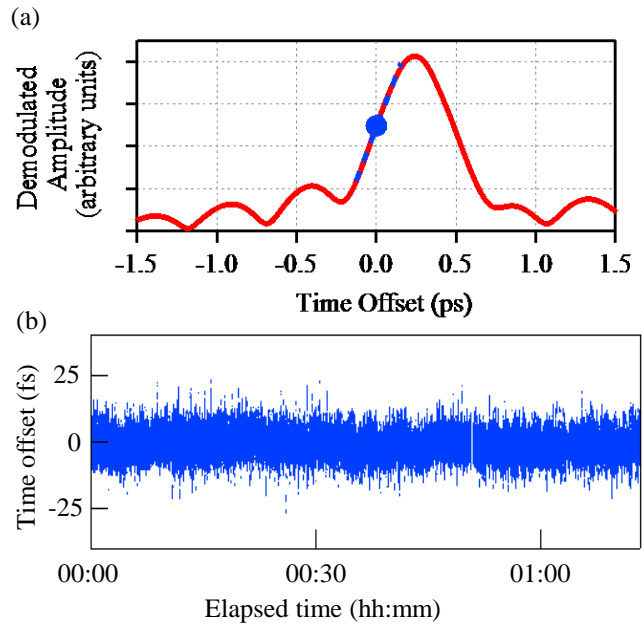

FIG 3 (Color) (a) Calibration of heterodyne out-of-loop measurement acquired by linearly sweeping $\Delta T_{AB}$ (through a software offset in the overall phaselocked loop of the time at site B). The blue dashed line indicates the linear range of response. (b) Typical measured out-of-loop time offset, $\Delta T$, after demodulation and appropriate calibration, over 4 km. While synchronized, the standard deviation is 2.4 fs.

## A. Synchronization over multiple days

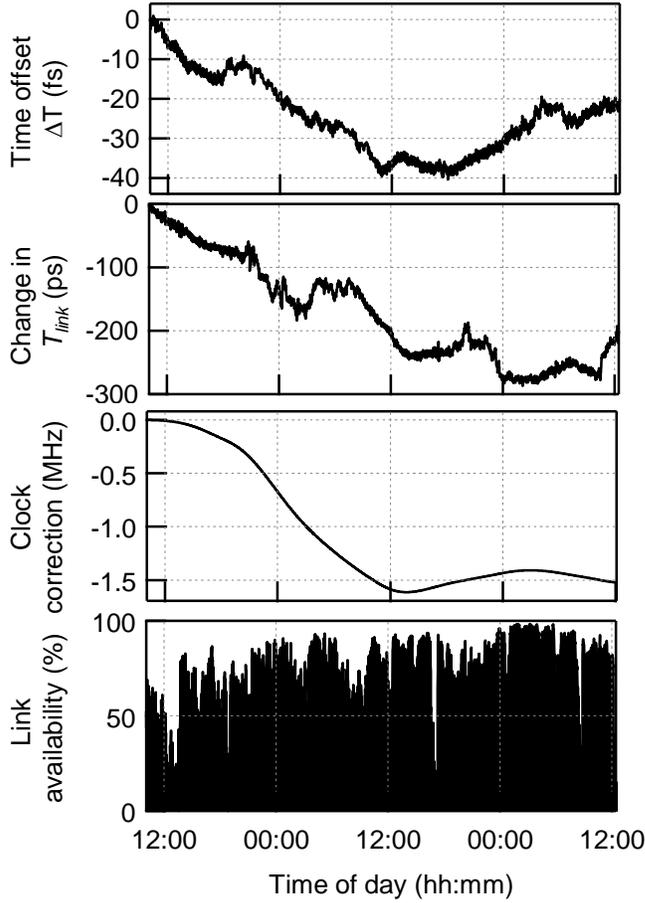

FIG 4. Synchronization data across 4 km over a 50 hour time period including, from top to bottom, the measured out-of-loop time offset $\Delta T$; the change in time of flight, $T_{link}$; the frequency correction applied to the timescale at site B to maintain synchronization; and the link availability. All data is filtered and downsampled from the 0.5 ms measurement period to 60 s.

Figure 4 summarizes an experiment where site A and B were synchronized for over 50 hours across a 4 km free-space link. The system ran without user intervention despite 4 °C rooftop laboratory temperature changes and ended with the arrival of a large snowstorm. (The system is able to operate through light snow and rain but not under heavy precipitation.)

The top panel plots the out-of-loop time offset as measured using the technique outlined in Fig. 3. These data are smoothed to 60 seconds. (An expanded view of the unsmoothed performance over short time periods was given in Fig. 3b.) The time-dependent offset is best analyzed by the timing deviation of these data, plotted in Fig. 1b, which is the uncertainty in the time offset as a function of averaging time [41]. From Fig. 1b, this uncertainty is below 1 fs out to 6500 s (1.8 hours), reaching a minimum of 225 attoseconds for a 10 s average. Therefore, we infer that the single spatial-mode link reciprocity over the 4 km air path is verified to below 70 nm at 10 s averaging and below 300 nm out to 6500 s. Fig. 4 shows that over the full 50 hour measurement, the time offset exhibits a larger 40 fs peak-to-peak wander. This time wander does not reflect a breakdown in reciprocity over the free-space link since a shorted link exhibits the same behavior. Instead, it reflects a weak temperature dependence of the system to the 4 °C laboratory temperature variations. Specifically, we attribute most of this wander to temperature-driven path length variations in the ~ 2 m of fiber patchcords that connect the two sites to the common reference plane and within the transceivers.

The second panel of Fig. 4 plots the change in $T_{link}$ over the measurement period. Its average value was ~13 µs, corresponding to the 3.942 km path distance. This time-of-flight variation corresponds to an 8.7 cm variation in optical path. The variation is driven by turbulence and building motion on short periods and by atmospheric temperature changes on longer periods. (Synnchronization under km-scale path variations are shown in the next section.)

The third panel of Fig. 4 plots the frequency correction that is applied to the 195.3 THz optical signal underlying the site B timescale. The effective time correction is given by the integral of this curve normalized by the nominal frequency of 195.3 THz and reaches 0.98 ms over the 50 hours reflecting the time wander between the two free-running cavity-stabilized lasers. One of the implicit byproducts of full synchronization is full syntonization, or "frequency lock". The residual frequency uncertainty between the sites is given by the modified Allan deviation, which is simply the timing deviation of Fig. 1b multiplied by $\sqrt{3}/t_{avg}$, where $t_{avg}$ is the averaging time. As shown in Fig. 5, this Allan deviation is consistent with the earlier 2 km comparison measurement of Ref. [35] despite the additional complexity of full time synchronization and longer distance. Moreover, it extends to longer averaging times reaching as low as $2\times10^{-19}$ beyond 10,000 sec.

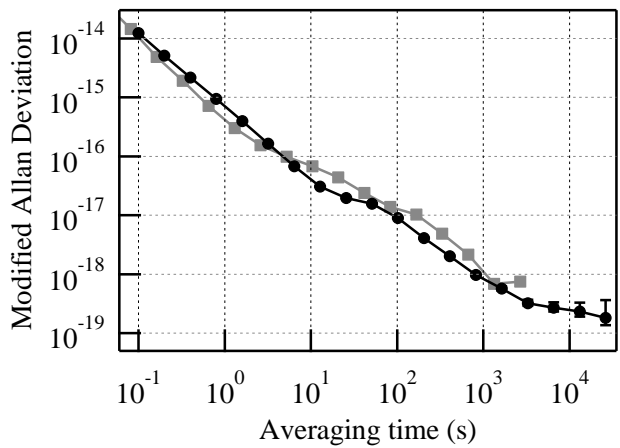

FIG 5. Modified Allan deviation for the corresponding frequency transfer from site A to site B (black trace). The fractional frequency uncertainty reaches $2\times10^{-19}$. Up to $10^3$ sec, the uncertainty in the syntonized frequencies matches the frequency-comparison measurement of Ref. [35] over 2 –km (gray trace) despite the additional complexity.

The bottom panel of Fig. 4 shows the "link availability" or the percentage of time per 60-second interval when sufficient optical power is transmitted over the link for two-way synchronization. Both the launched comb and communication/PRBS laser power are 2.5 mW, well below the eye safe limit. Atmospheric turbulence causes significant fluctuations in the received laser power in the single-mode fiber. The turbulence structure constant was $C_n^2 \approx 10^{-14}$ m$^{-2/3}$. With this moderate level of turbulence, the received power varies from 0 nW to ~200 nW, with a median value of 33 nW, which is compared to the detection threshold for the comb transfer of 2 nW, or ~78 photons per pulse (Fig. 6a). There are occasional "dropouts" when the received power is below threshold, leading to less than 100% link availability. These dropouts are typically below 10 ms in duration, as shown in Fig. 6b. During a dropout, the synchronization is not active and therefore these periods are excluded from the time offset data. Nevertheless, the cavity-stabilized lasers are well behaved so that the time offset at re-acquisition is typically below 6 fs. For systems that require a continuous output, a Kalman filter could be implemented. This is especially critical for less well behaved oscillators and long dropout durations but would be at the cost of significant added processing complexity.

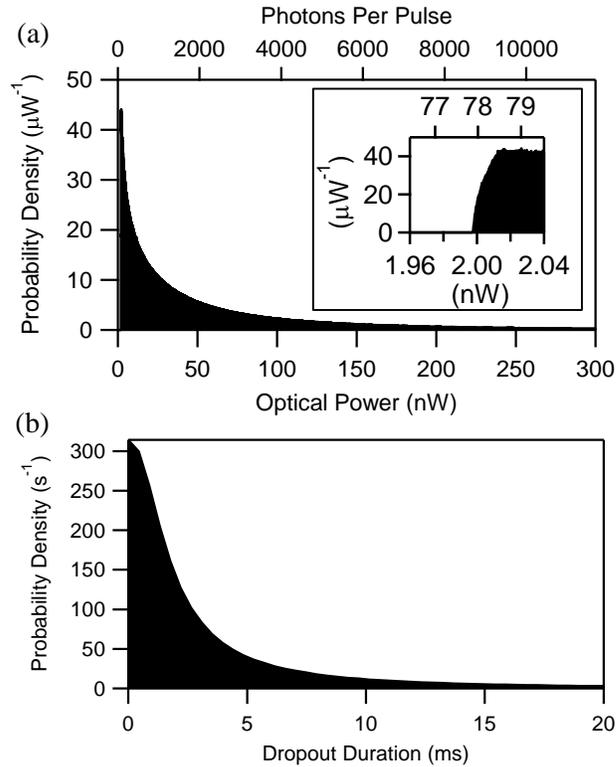

FIG 6. Turbulence-induced fades and power fluctuations. (a) Probability density of detected heterodyne signal optical power for the 50 hours of data shown in Fig. 4. Inset shows the ~ 2 nW threshold. (b) Probability density of fades versus duration. 90% of the fades are below 10 ms. Longer durations are typically due to a disruption of the beam from physical objects, re-alignment, or heavy precipitation, rather than turbulence.

### B. Synchronization maintained despite kilometer-scale length changes

The synchronization is robust against large changes in link distance. In Fig. 7, the out-of-loop time offset, $\Delta T$, is measured while the link distance is alternated between 1 m, 2 km and 4 km by manually adjusting the folding mirrors, as indicated in Fig. 7a. Each adjustment requires about 30 s. The system ran continuously during the link realignment, successfully re-synchronizing within tens of milliseconds of reacquisition of the light across the link. The overall time offset shows less than 2 fs wander with distance which is attributed to either a small systematic shift with distance or simply temperature variations within the laboratory (as mentioned in the previous section). Separate tests found negligible (< 1 fs) systematic shifts with received power.

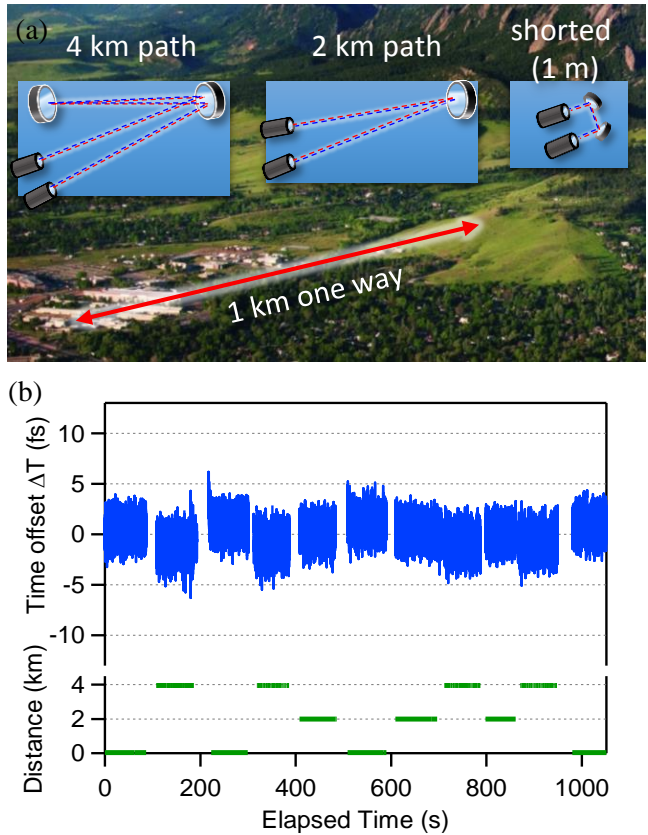

FIG 7. (Color) (a) The link traverses the NIST campus over 1 m, 2 km and 4 km distances with the latter achieved by a double pass between two flat mirrors. (b) The out-of-loop time offset, $\Delta T$, as the link distance was changed in real time.

## C. Optical pulse-per-second (PPS)

In conventional time systems, an rf PPS [24] provides unambiguous time markers. Here, we demonstrate femtosecond-level, unambiguous synchronization by generating analogous optical PPS signals. These optical PPS signals are easily generated by gating out a single pulse from the 200 MHz optical pulse train. At each site, the optical pulse train is fiber coupled to a Mach-Zehnder amplitude modulator (MZM) that is driven from a pulse generated by the local FPGA controller. Since this FPGA controller tracks the time associated with each optical pulses, it can gate every 200 millionth pulse (where we define our timescales such that the comb repetition rates are exactly 200 MHz). These pulses still carry the precision and accuracy of the synchronized timescales as they still consist of 150 fs long optical pulses. To verify unambiguous timing, each gated pulse is then photodetected and their arrival compared on a high bandwidth oscilloscope. To verify synchronicity, the common reference plane must be shifted by adjusting $\tau_{cal}$ from that of Fig. 4 to compensate for relative delays between photodetection and the oscilloscopes. Figure 8a shows an example of synchronization of 1 PPS signals to below ~100 ps, limited by the detector bandwidth. As with Fig. 7, this simultaneity is preserved across large path-length variations.

These data illustrate that the timing is unambiguous, but the uncertainty is limited by the rf bandwidths. As a more sensitive demonstration, we can spatially interfere the optical PPS from the two timescales. To do this, we construct a spatial interference fringe pattern by coupling the two optical outputs into free space and combining them at a slight angle onto an InGaAs focal plane array. A single PPS pair provides insufficient photons across the focal plane array so we increase the gate time to the MZM for a burst of pulses. Spatial interference fringes will be visible only when those bursts occur at the same time and only when the pulses within the burst overlap in time to well within their ~150 fs duration. The presence of the high-contrast spatial interference pattern indicates unambiguous, femtosecond-level synchronization between sites. Figure 8b shows such an interference pattern. The supplemental movie of Appendix A shows the appearance and disappearance of this spatial interference as synchronization is applied or disabled at the site B [42].

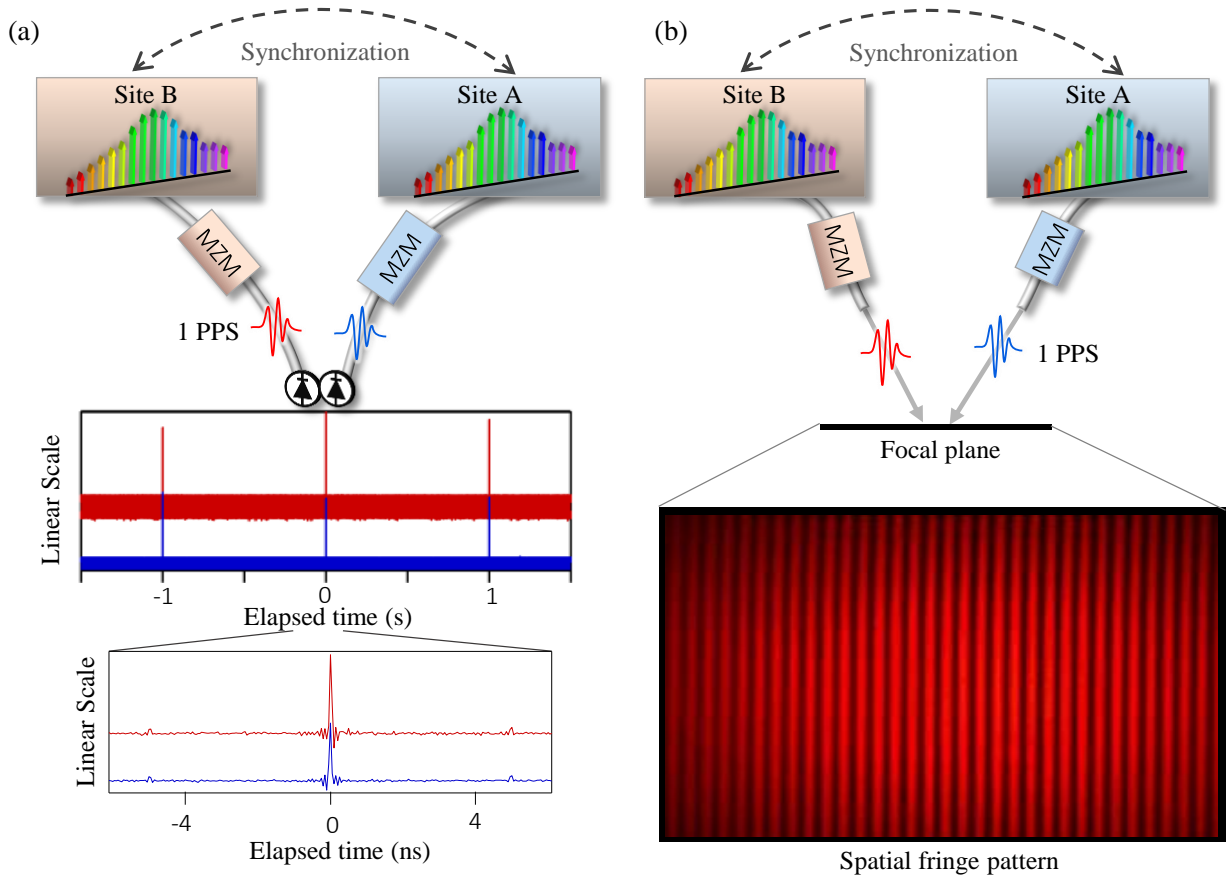

FIG 8. (Color, two column) Demonstration of synchronous optical pulse-per-second (PPS) outputs. (a) Synchronous optical PPS photo-detection at 8 GHz bandwidth. (b) Optical interference between selected pulse bursts measured through the tilt interference pattern on a focal plane array. The strong interference demonstrates that the pulses arrive within a time well below their correlation time of ~300 fs. (See also the supplementary video [41]). MZM: Mach-Zehnder modulator.

## V. Discussion

The results of the previous section demonstrate that the reciprocity of single-spatial mode optical links is sufficient to support femtosecond synchronization of distant optical timescales. Moreover, it is possible to achieve this synchronization in a complex, but robust implementation that can operate for days, over turbulent paths, and over path lengths of very different lengths.

In the system here, the two timescale are synchronized relative to each other to below 1 femtosecond for up to 1.8 hours. They are not stabilized to an absolute established timescale, although the master site A could be in principle. This low residual timing is nevertheless useful for a distributed passive or active sensing system or for navigation. For other applications it might be necessary to include an atomic clock at the master site A. For clock-based geodesy or relativity experiments, full atomic clocks are needed at each site for time comparisons. In that case, the residual timing noise associated with any comparison (or synchronization) between sites will be well below the absolute noise of the timescales.

Systematic time offsets with distance were below a few fs at 4 km, and no systematics were observed with variations in received optical power. There are however two important systematics. First, there will be temperature-induced path length changes in non-reciprocal optical paths within either the transceivers or in the verification. These effects can be suppressed by appropriate experimental design and by temperature control, down to tens of femtoseconds as shown here. Second, the width of the optical pulses is 100 fs to 1 ps long; the exact definition of the time associated with these pulses depends on how the end user "reads out" the arrival time of the pulse center at the reference plane, which will necessarily depend on the application. Again, this systematic will be on the order of tens of femtoseconds.

Rf-based two-way time-frequency transfer is much more developed and can operate over much longer ranges – including ground-to-space – and to moving platforms [24]. Here, our 4-km path is horizontal and therefore suffers equivalent turbulence to a longer vertical ground-to-satellite path, but longer distance operation will have higher transmission loss and path delay, $T_{link}$. The higher transmission loss will need to be offset by a reduced detection threshold, higher transmit powers, and improved free-space terminals, possibly including adaptive optics. The longer path delay can potentially cause a breakdown in the reciprocity condition, which assumes a "fixed" turbulence over the two-way measurement time of $1/\Delta f_r$. For $T_{link} \gg 1/\Delta f_r$, the short term turbulence-induced piston noise [38] will not be completely negligible but the long-term piston noise should nevertheless be cancelled via the two-way approach.

Moving platforms present at least two additional problems: point ahead issues and Doppler shifts. For transverse motion between platforms, the "point ahead" effect causes the two signals to traverse slightly different optical paths and therefore will cause a breakdown in reciprocity. As with the impact of a longer path delay, this effect will be strongest in a ground-to-space scenario. These effects have been analyzed recently by Wolf and coworkers, who find an increase in the timing noise over short times below a few seconds but excellent two-way cancellation over longer times [43]. The impact of Doppler shifts will require further study. To lowest order, the technique here is independent of Doppler shifts. However, the exact implementation here is not Doppler insensitive and future work is needed to optimize the system for insensitivity to Doppler shifts and to quantify any performance penalties.

## VI. Conclusions

We have demonstrated real-time time transfer and synchronization between remote optical timescales using two-way exchange of light over a reciprocal free-space link. We verify sub-femtosecond time synchronization out to 1.8 hours. The long-term wander over two days is 40 fs peak-to-peak, dominated by measurement uncertainty in the out-of-loop verification. The system was operated over a turbulent 4 km free-space path but we have found no fundamental limitations associated with distance. The single-mode free-space path is fully reciprocal to within our measurement uncertainty which reaches 70 nm at 10 second averaging. Provided sufficient received power is available (here equal to 78 photons per pulse), this approach should be scalable to much longer paths. The performance is a thousand times superior to rf based methods and should enable future networks of optical clocks/oscillators that are synchronized in real-time with sub-femtosecond stability.

### Appendix A: Supplemental movie

A supplemental movie is provided to show the spatial intensity of the overlapping PPS pulses from sites A and B, recorded with a SWIR camera (the focal plane array of Fig. 8) [42] during operation over a 4 km link. Spatial fringes appear and disappear as the synchronization is activated, de-activated, and re-activated.

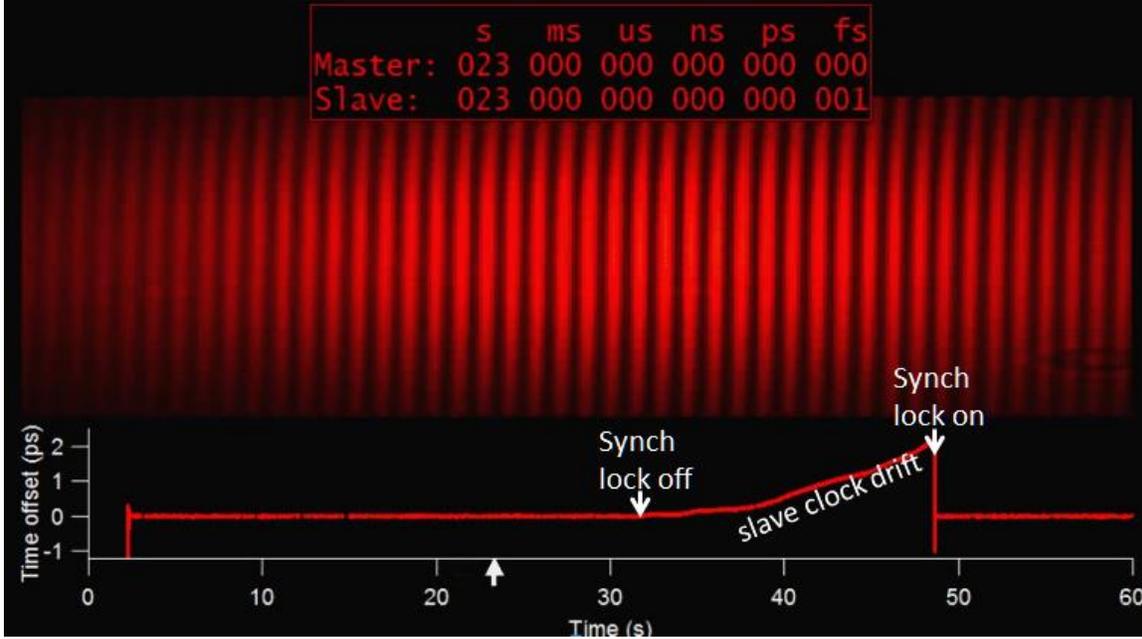

Fig. 9 (Color, two column) Snapshot from the supplementary movie.

### Appendix B: Derivation of the master synchronization equation.

We outline a derivation of the master synchronization equation. There are several factors that complicate the derivation. First, the derivation necessarily requires writing the time output of a clock versus time. In the time-frequency community, this notational challenge is sometimes handled by introducing the "$x$" variable for the clock time to write $x(t)$ or by viewing the clock output as a phase. The phase description is a useful one for this system as well, but will not be pursued here. Rather, we write the time marker from the $n$th comb pulse as $T(n) \equiv n f_r^{-1} + \tau(t)$ in terms of its repetition rate $f_r$ and time offset $\tau$, as a function of some "absolute" time, $t$ (which will not appear in the final synchronization equation). Second, there are multiple ambiguities that appear as "modulo" operations versus the comb pulse period, $1/f_r$, and the interferogram repetition period, $1/\Delta f_r$. These ambiguities must be appropriately handled for any absolute time comparisons between clocks.

The comb at site A produces a pulse train that is coherent with its cavity-stabilized cw laser,

$$E_A(t) = e^{i\phi_A} \sum_{n_A} e^{-i n_A \theta_0} A_A\left(t - n_A f_{r,A}^{-1} - \tau_A\right), \qquad (2)$$

where $\phi_A$ is an arbitrary phase, $\theta_0$ is the carrier-envelope offset phase, $n_A$ is a comb pulse index, $f_{r,A}$ is the repetition rate, $\tau_A$ is the time offset. In this form, it is clear the comb outputs pulses whose arrival time provides the time markers $T_A(n_A) \equiv n_A f_{r,A}^{-1} + \tau_A(t)$. Alternatively, the comb can be written as

$$E_A(t) = e^{i\phi_A} \sum_{k_A} \tilde{A}_{k,A} e^{-i2\pi\left(k_A f_{r,A} + f_{0,A}\right)(t - \tau_A)}, \tag{3}$$

where $k_A$ is the index of the comb tooth with complex amplitude $\tilde{A}_{k,A}$ at frequency $k_A f_{r,A} + f_{0,A}$, where the carrier-envelope offset frequency $f_{0,A} \equiv (2\pi)^{-1} \theta_0 f_{r,A}$. This second, equivalent form is useful in deriving the interferogram produced by the product of two combs below.

The other two combs have exactly the same form with variable subscripts "B" and "X" instead of "A". We will assume that the repetition rates are perfect, i.e. we attribute all of the time-varying clock error to $\tau$, which becomes a slowly varying function of time, $\tau(t)$. (In this case, slowly varying means slow on the time scale of $1/\Delta f_r$). We then denote the repetition rates as $f_{r,A} = f_{r,B} \equiv f_r$ and the difference in repetition rate between the site A comb and transfer combs as $\Delta f_r \equiv f_{r,X} - f_{r,A}$.

Linear optical sampling, as in Ref. 30, allows us to achieve femtosecond-level precision by recording the heterodyne signal between the pulse train from a remote comb and local oscillator comb, i.e. an interferogram. Interferograms are detected in three locations (given by the balanced detectors in Figure 1). These interferograms repeat with a period $1/\Delta f_r$ as the comb pulses walk through each other. The interferogram voltages are digitized by the local analog-to-digital converter (ADC), which is clocked at the local (site A or B) comb repetition rate. The interferograms digitized at site A have ADC sample times of $t_A = n_A f_r^{-1} + t_{0,A}$, where $t_{0,A}$ includes the time delay in detecting the comb pulse and any delays within the ADC itself. The interferogram digitized at the site B has sample time of $t_B = n_B f_r^{-1} + t_{0,B}$. One main purpose of the PRBS-TWTFT is to align these two ADC time bases by measuring their time offset $\Delta t_{ADC} \equiv t_A - t_B = \Delta n f_r^{-1} + t_{0,A} - t_{0,B}$, where $\Delta n \equiv n_A - n_B$. From the product of the comb electric fields (Eq. (3)) and including a low-pass filter, the digitized interferograms are the series

$$V_{A \to X}(t_A) \propto \sum_p A_{A \to X}\left(t_A - p\Delta f_r^{-1} - \frac{1}{\Delta f_r}\left[f_r(\tau_A + \delta\tau_A) - (f_r + \Delta f_r)(\tau_X + \delta\tau_X)\right]\right)$$

$$V_{B \to X}(t_A) \propto \sum_q A_{B \to X}\left(t_A - q\Delta f_r^{-1} - \frac{1}{\Delta f_r}\left[f_r(\tau_B + \delta\tau_B + T_{Link}) - (f_r + \Delta f_r)(\tau_X + \delta\tau'_X)\right]\right) \tag{4}$$

$$V_{X \to B}(t_B) \propto \sum_s A_{X \to B}\left(t_B - (-s\Delta f_r^{-1}) + \frac{1}{\Delta f_r}\left[f_r(\tau_B + \delta\tau'_B) - (f_r + \Delta f_r)(\tau_X + \delta\tau''_X + T_{Link})\right]\right)$$

after dropping any phase terms, where $p$, $q$, and $s$ are integers, $A_{A \to X}$ is the cross-correlation of the subscripted pulse envelopes again with an analogous definition for the two other series. For the first two interferograms, the transfer comb serves as the local oscillator, while for the third term, Site B's comb does, which causes the sign difference in the envelope terms. The $\delta\tau$ values represent the extra time delay in the transceivers associated with the comb reaching the relevant balanced detector. These are assumed constant with time and are removed through the calibration process. $T_{Link}$ is the time delay over the reciprocal single-mode path.

We extract the peak location of each interferogram after matched-filter processing to improve the signal to noise ratio. We then scale these peak locations by $\Delta f_r / f_r$ to find:

$$\Delta \tau_{A \to X}(p) = p\Delta f_r^{-1} + (\tau_A + \delta\tau_A) - \frac{(f_r + \Delta f_r)}{f_r}(\tau_X + \delta\tau_X)$$

$$\Delta \tau_{B \to X}(q) = q\Delta f_r^{-1} + (\tau_B + \delta\tau_B + T_{Link}) - \frac{(f_r + \Delta f_r)}{f_r}(\tau_X + \delta\tau'_X) \quad (5)$$

$$\Delta \tau_{X \to B}(s) = -s\Delta f_r^{-1} - (\tau_B + \delta\tau'_B) + \frac{(f_r + \Delta f_r)}{f_r}(\tau_X + \delta\tau''_X + T_{Link}) + \frac{\Delta f_r}{f_r}\Delta t_{ADC},$$

where we include the time offset between the sites' ADCs as $\Delta t_{ADC}$. Based on the PRBS-TWTFT, we align the $p$, $q$, and $s$ integers to compare the IGMs that are closest in time (to within $1/\Delta f_r$). The linear combination

$$\frac{1}{2}\left[\Delta\tau_{B \to X}(p) - \Delta\tau_{X \to B}(p) - 2\Delta\tau_{A \to X}(p)\right] = (\tau_A - \tau_B) + \left(\frac{\Delta f_r}{2f_r}\right)(T_{link} + \Delta t_{ADC}) + \delta\tau' \quad (6)$$

yields the slowly varying time offset between the sites (the first term) with additional contributions from the imperfect cancellation of the slowly varying link delay and ADC time offsets. The last term is the appropriate linear combination of the various $\delta\tau$ terms in Eq. (5), which are assumed constant.

We are interested in the time offset at the reference plane, which is defined as $\Delta T_{AB} = (\tau_A + \delta\tau_{ref,A}) - (\tau_B + \delta\tau_{ref,B})$, where $\delta\tau_{ref,A}$ is the fixed delay of the site A pulses to the site A reference plane and $\delta\tau_{ref,B}$ is similarly defined. (Here the two have the same reference plane so that we can verify synchronization.) We therefore rearrange (6) to find

$$\Delta T_{AB}(p) = \frac{1}{2}(\Delta\tau_{B \to X}(p) - \Delta\tau_{X \to B}(p)) - \Delta\tau_{A \to X}(p) + \tau_{cal} - \left(\frac{\Delta f_r}{2f_r}\right)(T_{link} + \Delta t_{ADC}) + \frac{\Delta n}{2f_r} \quad (7)$$

Or Eq. (1) in the main text where $\tau_{cal} = \delta\tau_{ref,A} - \delta\tau_{ref,B} - \delta\tau'$. $\tau_{cal}$ must be measured via a calibration step. Variations in $\delta\tau$ can lead to systematic time wander, as observed over the two day measurement. In this equation, we also explicitly add a term proportional to $\Delta n = n_A - n_B$ associated with the index of the pulses.

## Appendix C: Methods

The coarse two-way time transfer is accomplished via a phase modulated cw laser. At each site, the local field programmable gate array (FPGA) controller applies a phase modulation to a local DFB laser via an external Mach-Zehnder phase modulator. To enable coherent detection, the two DFBs lasers are frequency locked to an offset of 150 MHz by measurement of the incoming light from site A at the site B. At each site, the phase-modulated DFB laser is combined with the comb light through a wavelength division multiplexer (WDM). To implement the coarse two-way time transfer, [40] the site A first transmits a 80 μs long (~800 chips) Manchester-coded pseudo-random binary sequence (PRBS) phase-modulated laser signal at 100 ns chip length (~10 Mb/s signaling rate). When this is completed, site B transmits its own PRBS phase-modulated light signal across the link. Both sites use coherent detection to demodulation the PRBS signal and timestamp the arrival of the incoming signals according to their respective local ADC timebase. Because of this exchange of unique timestamps (64-bits), this measurement has for all practical matters infinite ambiguity range (5 ns * $2^{64}$). The difference yields $\Delta t_{ADC}$ and the sum yields $T_{link}$ unambiguously to within 40 ps uncertainty (one standard deviation) and resolves the pulse ambiguity, $\Delta n$.

For real-time communication, the same hardware is used. The coherent phase-modulated light operates in half-duplex mode using Manchester encoded binary phase shift keying (BPSK) at 10 Mbps and a protocol tolerant to link fades with low (10 µs) latency that only requires 350 µs for full bi-directional data and time transfer. Data integrity is ensured by a simple 10-bit cyclic redundancy error-detection code in each packet.

The combined comb and communication/PRBS light is launched across the 4 km path from single-mode fiber at the input of a free-space terminal. The free-space optical terminals use tip/tilt control to compensate for beam wander due to turbulence and building sway. A 5 mW beacon laser at either 1532.7 nm or 1542.9 nm, well separated from the other wavelengths, is polarization multiplexed with the comb and communication/PRBS light. The combined beams are then expanded in an off-axis, reflective parabolic telescope and launched over free space. At the receiver, the beam is collected by an identical terminal, and a dichroic then directs the beacon laser light to a quadrant detector, while the comb and communication/PRBS light are coupled into single-mode, polarization maintaining fiber which is then connected to the comb-based transceiver. The signals from the quadrant detector on each side are fed into an analog feedback system that controls the tip/tilt through an *x-y* galvanometric mirror pair, thereby centering the beacon laser and maximizing the comb and communication/PRBS light coupled into the single mode fiber.

### Acknowledgments

We acknowledge funding under the DARPA DSO PULSE program and helpful comments from Prem Kumar. We also acknowledge helpful comments from Marla Dowell, David Hume, and Andrew Ludlow, and technical assistance from Chris Cromer and Joe Thompson.